\input{aipcheck}

\documentclass[
    ,final            
  ,numberedheadings 
  ]
  {aipproc}

\layoutstyle{6x9}

\begin{document}

\title{Review of the experimental evidence on pentaquarks and critical discussion}

\author{Sonia Kabana}{
  address={Laboratory for High Energy Physics, University of Bern,
 Sidlerstrasse 5, 3012 Bern, Switzerland}
}

\begin{abstract}
We review and discuss the experimental evidence on 
 predicted baryonic states made by 4 quarks and one antiquark, called pentaquarks.
Theoretical and experimental advances in the last few years led
to the observation of  pentaquark candidates
by some experiments, however with relatively low individual significance.
Other experiments did not observed those candidates.
Furthermore, the masses of the $\theta^+$(1540) candidates exhibit a
large variation in different measurements. 
We discuss to which extend these contradicting
informations may lead to a consistent picture.
\end{abstract}

\maketitle

\section{ Introduction}

\vspace*{-0.3cm}

\noindent
Pentaquarks is a name devoted to describe baryons made by 4 quarks and one antiquark.
These states, predicted long time ago to exist \cite{pq_prediction,pra,diakonov_polyakov_petrov_1997},
  were searched for already in the 60'ies but few candidates
found have not been confirmed \cite{pdg_1986}.
Recent advances in theoretical \cite{diakonov_polyakov_petrov_1997}
 and experimental work \cite{nakano}
 led to
 a number of new  candidates in the last 2 years of searches 
\cite{nakano, clas_1, saphir,  clas_2, diana, hermes, neutrino, zeus, cosytof, camilleri,
 na49, na49_ximinus, h1, jamaica, graal_pentaquark2004, trento_graal}.
For recent reviews on pentaquarks see \cite{jaffe_review,pdg,reviews}.
 \noindent
The current theoretical description of pentaquarks
(e.g. \cite{diakonov_polyakov_petrov_1997, 
 jaffe, jaffe_2, kopel, octets_diakonov,ellis, stancu, aichelin_theta,stocker})
 does not lead to a unique picture
on the pentaquark existence and characteristics, reflecting the complexity of the subject. 
\noindent
In the following, we review and discuss the experimental observations of  pentaquark candidates,
as well as their lack of observation by some experiments.

\section{ Experimental evidence on pentaquark candidates}

\vspace*{-0.3cm}

\begin{figure}
  \includegraphics[height=.35\textheight]{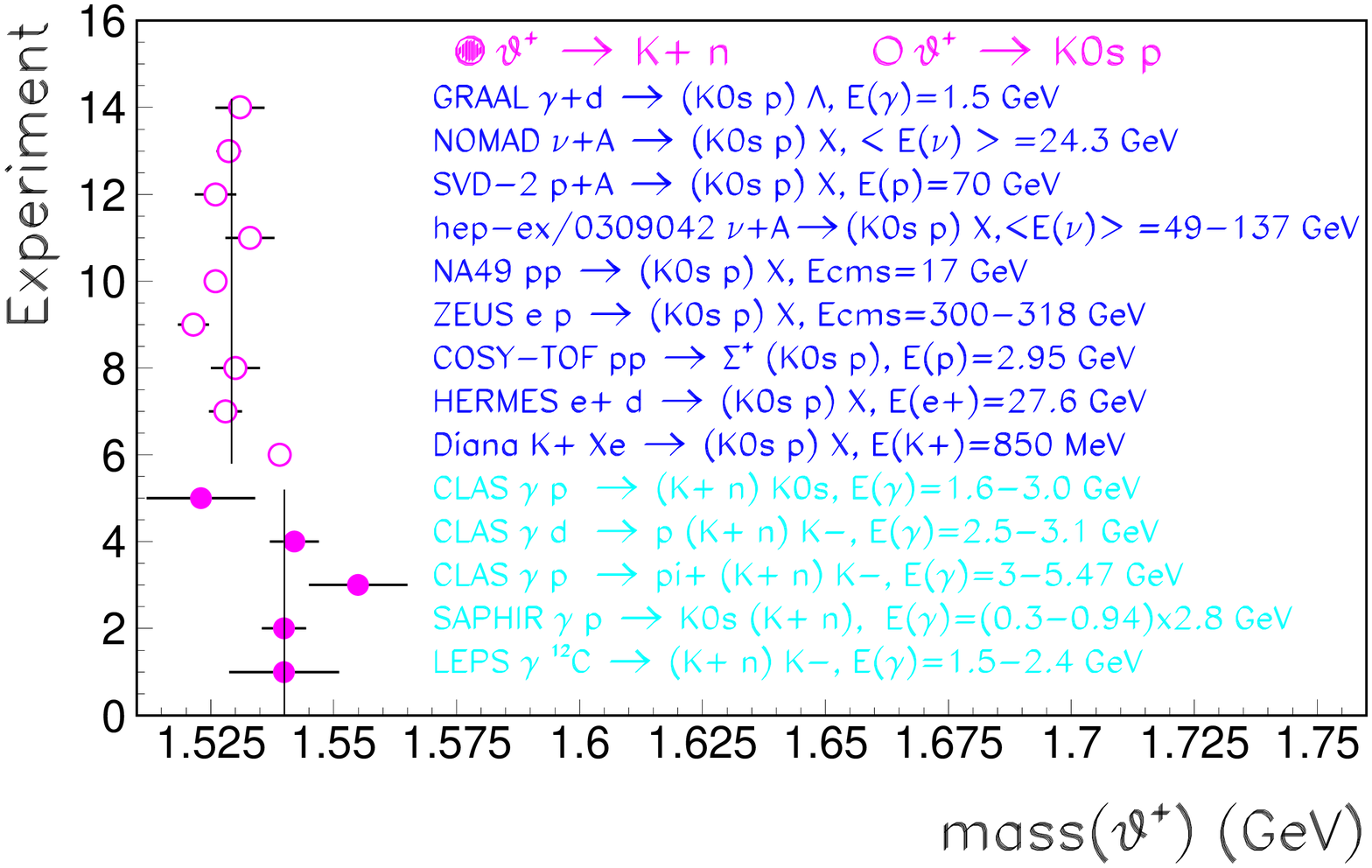}
\hspace*{-0.4cm}
  \includegraphics[height=.29\textheight]{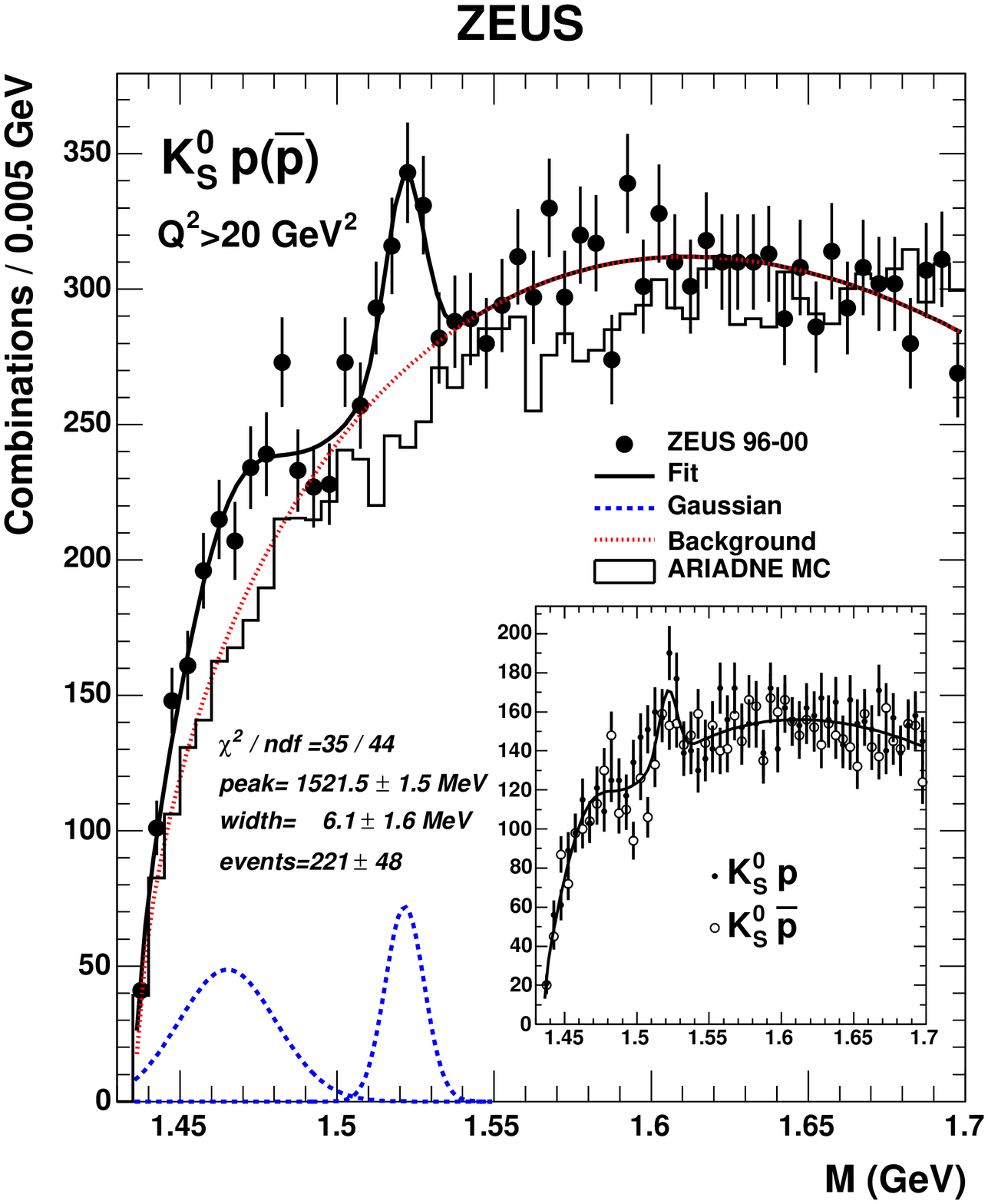}
\caption{Left: Compilation of measured masses of $\theta^+$ candidates.
Right: Zeus results on the $\theta^+$ candidate peak and its antiparticle.}
\label{theta_mass}
\end{figure}

\noindent
{\bf $\theta^+_{\overline{s}}$: $uudd \overline{s}$}
 \noindent
The first observation of a candidate for the $\theta^+$ pentaquark has been
reported by the LEPS collaboration \cite{nakano} in
reactions $\gamma + A $ with $\gamma$ energy 1.5-2.4 GeV 
and in the decay channel $n K^+$.
Recent preliminary analysis of new data taken recently by LEPS lead to a confirmation
of the seen peak with about 90 entries in the peak above background, as compared
to 19 measured previously \cite{leps_pentaquark2004}.
\noindent
This first observation were followed by a number of experiments which have seen
the $\theta^+$ candidate peak 
\cite{clas_1, saphir, clas_2, diana, hermes, neutrino, zeus, cosytof, camilleri, na49}.
Figure
 \ref{theta_mass}, left, shows the masses of all $\theta^+$ candidate peaks measured.
The $\theta^+$ peak has been observed in two decay channels.
The open points correspond to the decay channel $K^0_s p$, while the
closed points  to the decay channel  $K^+ n$.
The
 candidate $\theta^+$ peak is seen in different reactions namely of 
$\gamma+A$, $\nu +A$, $p+p$, $K +Xe$, $e +d$, $e+p$, $K +Xe$.
All of these reactions involved at least a baryon in the initial state.
The energies are small (few GeV range) for all $\gamma+A$ reactions
and vary for the rest up to $\sqrt{s}$=300 GeV for $e +p$.
In the experiments measuring the decay channel  $K^+ n$  the neutron
was not directly measured.
Even though the $\theta^+$ candidate peak has been observed by several experiments,
  the individually achieved statistical significance
of the signal is mostly not large. The largest significance
was  $S/ \sqrt{B}$ = 7.8 $\pm$ 1 
\cite{clas_2}.
\\
A remarkable observation has been made by the CLAS  collaboration
in the same publication \cite{clas_2},
namely they  observed that the $\theta^+$ candidate seem to be preferably
produced through the decay of a possible new narrow resonance $N^0(2400)$.
 A preliminary analysis of CLAS \cite{bata_pentaquarks2004} showed also a second peak 
 in the invariant mass ($K^+ n$)
 at 1573 $\pm$ 5 MeV with a significance of about 6 $\sigma$.
The second peak is a candidate for an excited $\theta^+$ state which is expected 
to exist with about $\sim$ 50 MeV higher mass than the ground state, in
agreement with the observation.
A preliminary cross section estimate gives 5-12 nb for the low mass peak
and 8-18  nb for the high mass peak.
\noindent
Cross sections have been reported also from 
the COSY-TOF  collaboration
\cite{cosytof}
(proton beam 2.95 GeV on protons)
which observed a $\theta^+$ peak in the invariant mass $p K^0_s$.
They measure a cross section of 0.4 $\pm$ 0.1 $\pm$ 0.1 (syst) $\mu b$
which is in rough agreement with predictions of
0.1-1 $\mu b$
for p+p, p+n near threshold.
The ZEUS  collaboration 
  \cite{zeus}
  ($e^+p$ $\sqrt{s}$=300-318 GeV)
is the only experiment which
 observed for the first time the $\overline{ \theta}^-$
state decaying in $\overline{p} K^0_s$ (fig. \ref{theta_mass}, right).

\noindent
Most experiments measure a $\theta^+$ width consistent with the experimental resolution,
while Zeus and Hermes give a measurement of width somewhat larger than their resolution.
A measurement with a much improved resolution would be important.
Non-observation of $\theta^+$ in previous experiments lead to an estimate
of its width to be of the order of 1 MeV or less \cite{1mev_arndt}.
This limit would gain in significance, once the lack of observation of the $\theta^+$
peak by several experiments will be better understood.

\noindent
{\bf Study of the $\theta^+$ mass variation}

\noindent
Figure \ref{theta_mass}, left, as previously mentioned shows a compilation of the 
masses of $\theta^+$ candidate peaks observed by several experiments.
The statistical and systematic errors (when given) have been added in quadrature.
For GRAAL 
we assume an error of 5 MeV as no error has been given in \cite{graal_pentaquark2004}.
For the two preliminary peaks of CLAS we assume the systematic error
of 10 MeV quoted previously by CLAS.
The lines indicate the mean value of the mass among 
the $\theta^+ \rightarrow p K^0_s$
and the
$\theta^+ \rightarrow n K^+$
observations.

\noindent
It appears that the 
mass of $\theta^+$
from $\theta^+ \rightarrow n K^+$
observations
is systematically higher than the one
from 
$\theta^+ \rightarrow p K^0_s$
observations.
This may be related to the special corrections needed
for the Fermi motion and/or to details of the
analysis with missing mass instead of direct measurement 
of the decay products.

\noindent
All observations together give a mean mass of 
1.533 $\pm$ 0.023 GeV and they
deviate from their mean with a
 $\chi^2/DOF$
of 3.92.
The $\chi^2/DOF$ for 
the deviation of the $\theta^+ \rightarrow p K^0_s$
observations
from their mean of 1.529 $\pm$ 0.011 GeV
is 
3.76.
The $\chi^2/DOF$ for 
the deviation of the $\theta^+ \rightarrow n K^+$
observations
from their mean of 1.540 $\pm$ 0.020 GeV
is 0.94.

\noindent
The bad $\chi^2/DOF$ for the $\theta^+ \rightarrow p K^0_s$
observations
maybe due to an underestimation of the systematic errors.
In particular in some cases no systematic errors are given,
sometimes because the results are preliminary.
If we add a systematic error of 0.5\% of the measured mass (therefore
of about 8 MeV) on all
measurements for which  no systematic error 
was given by the experiments,
we arrive to a
$\chi^2/DOF$ for the $\theta^+ \rightarrow p K^0_s$
observations
of 0.95
and a mean mass of 1.529 $\pm$ 0.022 GeV.
The $\chi^2/DOF$ for the $\theta^+ \rightarrow n K^+$
observations
almost don't change by this,
(mean mass = 1.540 $\pm$ 0.022 GeV, $\chi^2/DOF$=0.91),
because the experiments mostly give the systematic errors
for this decay channel.
All observations together give then a mean mass of 
1.533 $\pm$ 0.031 GeV and they
deviate from their mean with a
 $\chi^2/DOF$
of 2.1, reflecting mainly the difference of masses between
the two considered decay channels.
It is important to understand the origin of this discrepancy.
This
 can be studied measuring $\theta^+ \rightarrow K^+ n$ in experiments
with direct detection of the neutron or the antineutron for the $\overline{ \theta^-}$
 like PHENIX and GRAAL.

\vspace*{0.1cm}

\noindent
{\bf $\theta^{++}$ }
\noindent
A preliminary peak is quoted by CLAS \cite{bata_pentaquarks2004}
 for the candidate $\theta^{++} \rightarrow p K^+$
 produced in the reaction 
$ \gamma p \rightarrow \theta^{++} K^-
		\rightarrow p K^+ K^-$
 at 1579 $\pm$ 5 MeV.
A previous peak observed by CLAS in the invariant mass $ p K^+$
has been dismissed as due to $\phi$ and hyperon resonance reflexion
\cite{clas_thetapp_1}.
\noindent
The STAR collaboration quoted a preliminary peak in the $p K^+$ and 
 $\overline{p} K^-$ invariant masses at 1.530 GeV
in d+Au collisions at $\sqrt{s}$=200 GeV \cite{star_thetapp}.

\vspace*{0.1cm} 
\noindent
{\bf $\Xi$, $N^0$}
\noindent
The NA49 experiment has observed in p+p reactions
at $\sqrt{s}$=17 GeV 
the pentaquark candidates 
$\Xi^{--}(1862 \pm 2  MeV) \rightarrow \Xi^- \pi^-$,
the
$\Xi^{0}(1864 \pm 5  MeV) \rightarrow \Xi^- \pi^+$
and their antiparticles  \cite{na49}.
They measure a width consistent with their resolution of about  18 MeV.
They also observe preliminary results of the decay
$\Xi^-(1850) \rightarrow \Xi^0(1530) \pi^-$
with simarly narrow width as the other candidates \cite{na49_ximinus}.

\noindent
The experiment STAR has shown preliminary results on a 
$N^0$ ($udsd \overline{s}$)
 or $\Xi$ ($uds s \overline{d}$) I=1/2 candidate \cite{jamaica}.
STAR uses minimum bias Au+Au collisions at $\sqrt{s}$=200 geV
and  observes a peak in the decay channel
	$\Lambda K^0_s$ at a mass 1734 $\pm$ 0.5 (stat) $\pm$ 5 (syst) MeV
with width consistent with the experimental resolution of about  6 MeV
and $S/ \sqrt{B}$ between 3 and 6 depending on the method used
\cite{jamaica}.

\noindent
The GRAAL experiment has shown preliminary results on 
two narrow  $N^0$ candidates.
One candidate is observed at a mass of 1670 MeV in  the invariant mass of
$ \eta n$ from the reaction
$ \gamma d \rightarrow \eta n X$.
 The neutron has been directly detected.
The other is observed at a mass of 1727 MeV
 in the invariant masses of 
$\Lambda K^0_s$  as well as 
in the invariant masses of
$\Sigma^- K^+$
at the same mass and with the same width
\cite{trento_graal}.
The second reaction allow to establish the strange quark content  and
therefore to exclude the $\Xi$ hypothesis.
The difference of 7 MeV between the STAR and GRAAL 
measured masses of 1727 and 1734 MeV, should be compared to the
systematic errors. STAR quotes a systematic error of 5 MeV
while GRAAL quotes no systematic error.

\noindent
The mass of the peaks  at 1670 and at  (1727,1734) MeV  is in  good agreement with the
$N$ masses 
  suggested by Arndt et al \cite{arndt0312126}. In this paper
a modified Partial Wave analysis allows to search for narrow
states and presents two candidate $N$ masses, 1680 and/or 1730 MeV
with width below 30 MeV.

\vspace{0.1cm} 
\noindent 
{\bf $\theta^0_{\overline{c}}$}
\noindent
The H1  collaboration 
at DESY
  used $e^- p$ collisions at $\sqrt{s}$=300 and 320 GeV 
 and
have observed a peak in the invariant masses
$ D^{*-} p$
and
$ D^{*+} \overline{p}$
at a mass 3099 $\pm$ 3 (stat) $\pm$ 5 (syst) MeV
and width of 12 $\pm$ 3 MeV  \cite{h1}.
This peak is a candidate for the state $\theta^0_{\overline{c}} $ = $uudd \overline{c}$
and is the first charmed pentaquark candidate seen.

\vspace{0.2cm} 
\noindent
{\bf Lack of observation of pentaquark candidates}

\noindent
Several experiments have reported preliminary or final results on the non-observation
of pentaquarks e.g.
 $e^+e^-$: Babar, Belle, Bes, LEP experiments, 
$p \overline{p}$: CDF, D0
 $pA$:E690,
$\gamma p$: FOCUS,
 $pA$: HERA-B,
 $ep$: Zeus (for the $\theta^0_c$) 
$ \mu^+ \ ^6 Li D$: COMPASS,
Hadronic Z decays: LEP,
$\pi$, $K$, p on A: HyperCP,
$\gamma \gamma$: L3,
$\pi, p, \Sigma$ on p: SELEX,
$pA$: SPHINX,
$\Sigma^- A$: WA89
$K^+ p$: LASS,
\cite{non}.
\noindent
 HERMES has reported the non-observation of a $\theta^{++}$ candidate peak 
 in the $p K^+$ invariant masses \cite{hermes}.
No other experiment
has observed the candidates for the $\Xi$ and the $\theta_c$ pentaquarks
seen by NA49 and H1. 
Especially Zeus has searched for the $\theta_c$ under similar
conditions as H1 and with similar statistics, without observing a peak \cite{zeus_sardegna}.
Many of the experiments reporting non observation of pentaquarks
have a very high statistics and good mass resolution.

\noindent
It has been  argued that the non-observation of pentaquark
states in the above
experiments could be due to an additional strong suppression factor
for pentaquark production
in $e^+ e^-$ collisions, as well as in B decays 
which is lifted in reactions like $\gamma A$
in which
a baryon  is present in the initial state
\cite{karliner}.
The constituents of the $\theta^+$ are already present in the initial
state of e.g.  low energy photoproduction experiments, while
in other experiments baryon number and strangeness must be created from gluons \cite{karliner}.
It is important to try to assess the expected cross sections.
\\
The non observation of pentaquarks
 in high energy interactions of hadrons (CDF ($p \overline{p}$), E690 (pA) etc)
can be a consequence of the decrease of the pentaquark cross section
with increasing energy \cite{0406043,titov}.
This depends however on the kinematic region considered, and 
it is suggested to look for pentaquarks in the central rapidity region  \cite{0406043,titov}.
\\
In addition, if the $\theta^+$
is produced preferably through the decay of a new resonance 
$N^0(2400) \rightarrow \theta^+ K^-$ 
as suggested by CLAS and NA49 and as discussed in \cite{karliner,azimov_n2400},
neglecting this aspect maybe a further cause of its non-observation in some experiments.
\\
Some authors pointed out the importance to 
exclude kinematic reflexions as reason behind the $\theta^+$ peak
\cite{dzierba}.
This known source of systematic errors is under investigation by the experiments
which observe pentaquark candidates.

\noindent
It is clear that a higher statistic is desirable in order to confirm
the pentaquark observations reported so far.
New data taken in 2004 and planned to be taken in 2005 will
lead to enhancements in statistics of experiments up to a factor of 15
allowing to test the statistical significance and make more systematic studies.
Experiments searching for pentaquarks should  test also the production mechanisms 
proposed in the literature e.g. the $\theta^+$ production
through the $N^0(2400)$ decay.
For example
 Phenix could  search for the final state
$ \overline{ \theta^-} K^+ $
or 
$ \overline{ \theta^-} K^0_s $
demanding the invariant mass of $ \overline{ \theta^-} K^0_s $ and
$ \overline{ \theta^-} K^+ $
to be in the range 2.3 to 2.5 GeV,
and study the option to trigger online on this channel.

 \vspace*{-0.2cm}

\section{Summary and conclusions}

 \vspace*{-0.2cm}

\noindent
Recent theoretical and experimental advances led to the observation of
candidates for a number of pentaquarks states.
In particular candidate signals have been observed for the
$\theta^+(1533)$, $\theta^{++}$(1530/1579),
$\theta^0_{\overline{c}}$(3099),
$\Xi^{--}(1862)$,
$\Xi^{0}(1864)$,
$\Xi^{-}(1850)$,
$N/\Xi^0(1734/1727)$,
$N^0(2400)$,
$N^0(1670)$ states
as well as a possible excited $\theta^+_{\overline{s}}(1573)$ state.
These
 observations are promising, however despite the large number of e.g. the
$\theta^+_{\overline{s}}$ observations, they all suffer
 from a low individual statistical significance.
A 
much higher statistics is needed to support and solidify the existing evidence.
Several other high statistics experiments have reported lack of observation of those candidates.
The
 inconsistency among experiments waits to be clarified through
 high statistics measurements of pentaquark candidates, 
their characteristics (cross sections, quantum numbers)
  and  upper limits in the case
of non-observation. 
Furthermore, systematic studies are needed as well as advanced theoretical
understanding of the observations, in particular of the narrow width and the
production mechanism of the observed candidates as well
as the possible reasons behind their non-observations by some experiments.
Combined theoretical and experimental efforts
should be able to answer soon the question if pentaquarks exist.

\end{document}